\documentclass[a4paper,11pt]{article}
\usepackage[dvipdfmx]{graphicx}
\usepackage[hang,small,bf]{caption}
\usepackage[subrefformat=parens]{subcaption}
\usepackage{bm}
\usepackage{pos}

\title{Dilepton production rate near the critical temperature of color superconductivity%
\\ \vspace*{-70mm}\hspace{12.3cm} \small{\textrm{J-PARC-TH-0251}} \vspace*{68mm}}
\ShortTitle{Dilepton production rate near the critical temperature of CSC}

\author*[a]{Toru Nishimura}
\author[a,b]{Masakiyo Kitazawa}
\author[c]{Teiji Kunihiro}

\affiliation[a]{Department of Physics, Osaka University, \\
560-0043, Toyonaka, Osaka, Japan}

\affiliation[b]{J-PARC Branch, KEK Theory Center, Institute of Particle and Nuclear Studies, KEK, 203-1, \\
319-1106, Shirakata, Tokai, Ibaraki, Japan}

\affiliation[c]{Yukawa Institute for Theoretical Physics, Kyoto University, \\
606-8502, Kyoto, Japan}

\emailAdd{nishimura@kern.phys.sci.osaka-u.ac.jp}
\emailAdd{kitazawa@phys.sci.osaka-u.ac.jp}
\emailAdd{kunihiro@yukawa.kyoto-u.ac.jp}

\abstract{
  We investigate modification of the dilepton production rate
  by the diquark fluctuations that form well-developed collective modes
  near the critical temperature of color superconductivity.
  Through the analysis of the photon self-energy called
  the Aslamasov-Larkin, Maki-Thompson and density of states terms
  in the theory of metalic superconductivity, 
  it is shown that the collective mode in the diquark channel
  affects the photon self-energy significantly 
  and thereby gives rise to an anomalous enhacement of 
  the dilepton production rate in the low invariant-mass region.
}

\FullConference{%
	The International conference on Critical Point and Onset of Deconfinement - CPOD2021\\
  	15 – 19 March 2021\\
 	Online - zoom
 }

\begin{document}
\maketitle

\section{Introduction}

Experimental programs in relativistic heavy-ion collisions (HIC) 
such as the beam-energy scan program at RHIC, HADES and NA61/SHINE,
as well as the future plans at FAIR, NICA and J-PARC-HI,
are aimed at revealing rich physics 
in high baryon-density matter at finite temperature.
In this report,
we theoretically explore  the possibility to observe
precursory phenomena of 
the color superconductivity (CSC) \cite{Alford:2007xm}
in these experiments through the analysis of the dilepton production rate,
on the basis of the observation \cite{Kitazawa:2001ft,Kitazawa:2005vr}
that the diquark fluctuations 
are developed in the temperature higher than but
near the critical temperature $T_c$ of the CSC. 
We show that the diquark fluctuations modify 
the photon self-energy and thereby affect the dilepton production 
rates near $T_c$ by extending the theory of the paraconductivity
in metals~\cite{Kunihiro:2007bx}.

\begin{figure}[bp]
\centering
\includegraphics[scale=0.27]{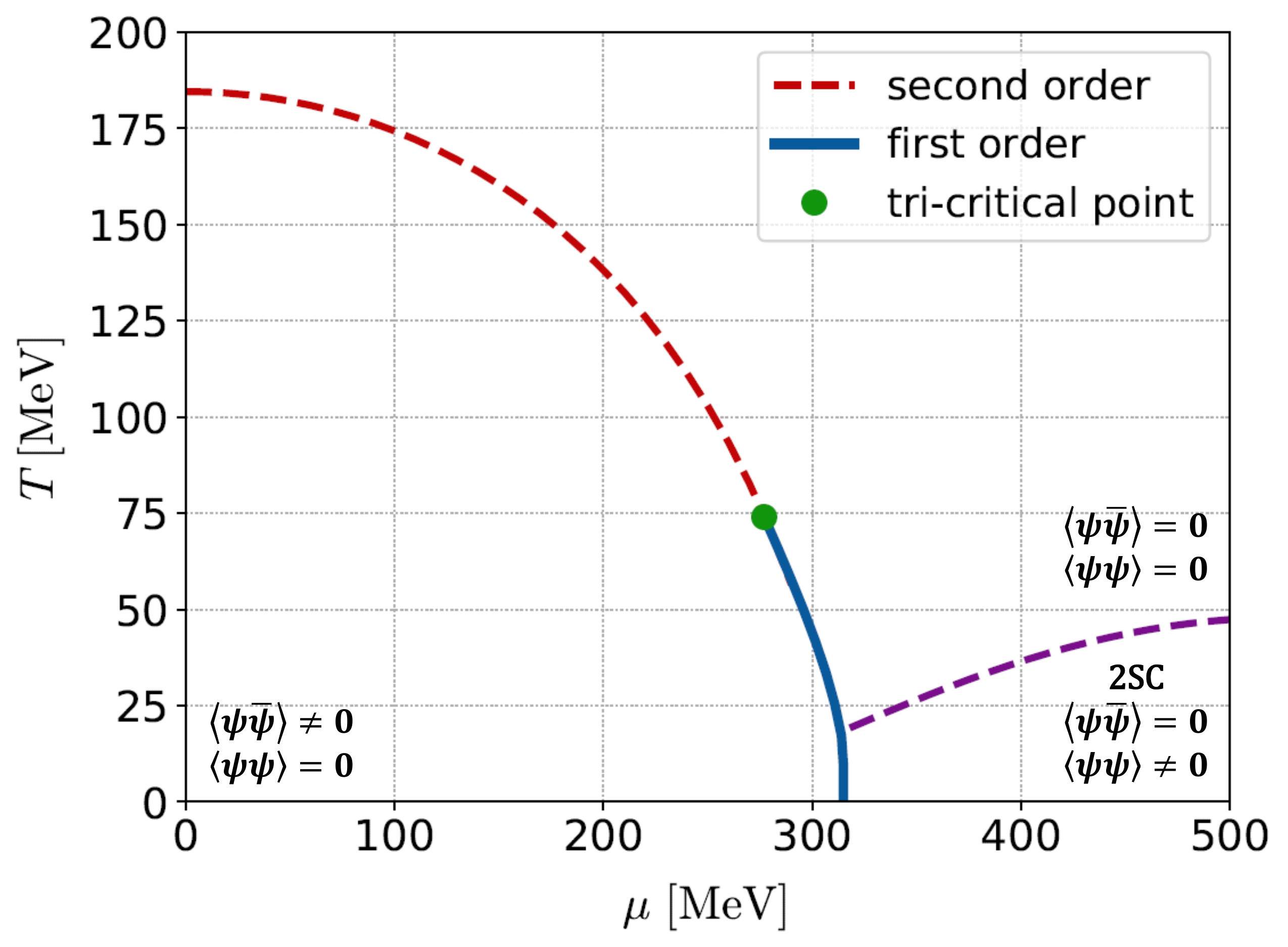}
\caption{
  The phase diagram obtained in the massless 2-flavor 
  NJL model Eq.~(\ref{eq_lagrangian}).
  The dashed lines show the second-order phase transitions.
}
\label{fig_phase}
\end{figure}

\section{Model and phase diagram}

We focus on the 2-flavor color-superconductivity (2SC), 
which is expected to be realized at relatively low densities, 
and employ the massless 2-flavor and 3-color NJL model~\cite{Kitazawa:2001ft,Kitazawa:2005vr},
\begin {align}
\mathcal{L} = \bar{\psi} i \gamma_{\mu} \partial^{\mu} \psi 
\ + \ G_S [(\bar{\psi} \psi)^2 + (\bar{\psi} i \gamma_5 \vec{\tau} \psi)^2] 
\ + \ G_C (\bar{\psi} i \gamma_5 \tau_2 \lambda_A \psi^C)(\bar{\psi}^C i \gamma_5 \tau_2 \lambda_A \psi), \label{eq_lagrangian}
\end {align}
where the second and third terms represent the $qq$ 
and  $q \bar{q}$ interactions, respectively, and
$\psi^C (x) \equiv C \bar{\psi}^T (x)$ with $C = i \gamma_2 \gamma_0$.
$\tau_2$ and $\gamma_A$ $(A=2,5,7)$ are the antisymmetric components 
of the Pauli and Gell-mann matrices for the flavor $SU(2)_f$ 
and color $SU(3)_c$, respectively.  
The scalar coupling constant $G_S=5.01 \rm{MeV^{-2}}$ and the
three-momentum cutoff $\Lambda=650$~MeV are
determined so as to reproduce the pion decay constant 
$f_{\pi}=93 \rm{MeV}$ and the chiral condensate 
$\langle \bar{\psi} \psi \rangle = (-250\rm{MeV})^3$ 
in the chiral limit. 
The diquark coupling constant is set to $G_C = 0.6G_S$.
We show the phase diagram obtained in the mean-field
approximation (MFA) with the mean fields
$\langle \bar{\psi} \psi \rangle$ and 
$\langle \bar{\psi}^C i \gamma_5 \tau_2 \lambda_A \psi \rangle$
in Fig.~\ref{fig_phase}.
The 2SC phase is realized at low temperature and high density region.
In the following, we focus on the medium near but above the
critical temperature of 2SC.

\section{Photon self-energy due to the diquark fluctuations}

The dilepton production rate is given in terms of the 
retarded photon self-energy $\Pi^{R \mu\nu}(k)$ as 
\begin {align}
\frac{d^4\Gamma}{d^4k} = \frac{\alpha}{12\pi^4} 
\frac{1}{k^2} \frac{1}{e^{\beta\omega}-1} g_{\mu\nu}  
{\rm Im} \Pi^{R \mu\nu} (k) ,\label{eq_RATE_kom}
\end {align}
where $k=(\bm{k}, \omega)$ is the four momentum
of the photon and $\alpha$ is the fine structure constant. 

In Refs.~\cite{Kitazawa:2001ft,Kitazawa:2005vr},
it has been pointed out that 
the diquark fluctuations 
 develop the collectivity at temperatures above but 
near the critical temperature of the CSC.
In the present study we investigate the modification of the 
photon self-energy due to the diquark fluctuations.
The photon self-energy is derived so that it satisfies
the Ward-Takahashi (WT) identity
$k_\mu \Pi^{R \mu\nu} (k)=0$ utilizing the thermodynamic potential.

\subsection{Diquark fluctuation mode}

\begin{figure}[tbp]
\centering
\includegraphics[scale=0.26]{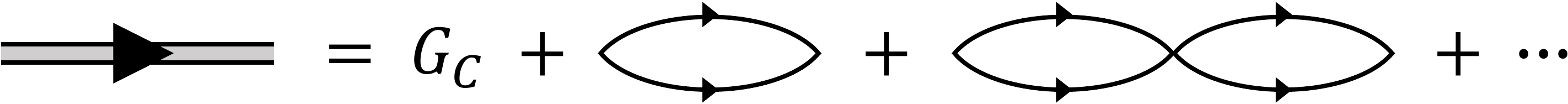}
\caption{Diagrammatic representation for the propagator Eq.~(\ref{eq_softmode}).}
\label{fig_softmode}
\end{figure}

The propagator of 
the diquark fluctuations in the random-phase approximation
(Fig.~\ref{fig_softmode}) is given by
\begin {align}
  \Xi(\bm{q}, i\nu_n) = \frac{G_C}{1+G_C\mathcal{Q}(\bm{q}, i\nu_n)},
  \label{eq_softmode} 
\end{align}
where $\mathcal{Q}(\bm{q}, i\nu_n)$ is the one-loop quark-quark correlation
\begin{align}
  \mathcal{Q} (\bm{q}, i\nu_n) = -2 N_f(N_c-1) T\sum_m
  \int \frac{d^3p}{(2\pi)^3} {\rm Tr} [C \Gamma \mathcal{G}_0 
    (\bm{q-p}, i\nu_n-i\omega_m) \Gamma C \mathcal{G}_0^T(\bm{p}, i\omega_m)],
\end{align}
$\omega_m$ ($\nu_n$) is the Matsubara frequency for fermions (bosons),
${\rm Tr}$ is the trace over the Dirac indices and 
$\mathcal{G}_0(\bm{p}, i\omega_m) = 1/[(i\omega_m + \mu)\gamma_0 - \bm{q} \cdot \bm{\gamma}]$ 
is  the free quark propagator.
By taking the analytic continuation 
$\Xi^R (\bm{q}, \omega) = \Xi (\bm{q}, i\nu_n \rightarrow \omega+i\eta)$
one obtains the retarded propagator.
We remark that
$[\Xi^{R} (\bm{0}, 0)]^{-1} = 0$ is satisfied at $T=T_c$ determined
by the MFA, 
which is nothing but the Thouless criterion for determining the critical
temperature of the second-order phase transition.
The Thouless criterion shows that the diquark propagator
$\Xi^{R} (\bm{q}, \omega)$ has a pole at the origin at $T=T_c$,
and hence the diquark fluctuations have the properties of
the soft mode~\cite{Kitazawa:2001ft,Kitazawa:2005vr}.

\subsection{Photon self-energy}
\begin{figure}[tbp]
\centering
\includegraphics[scale=0.33]{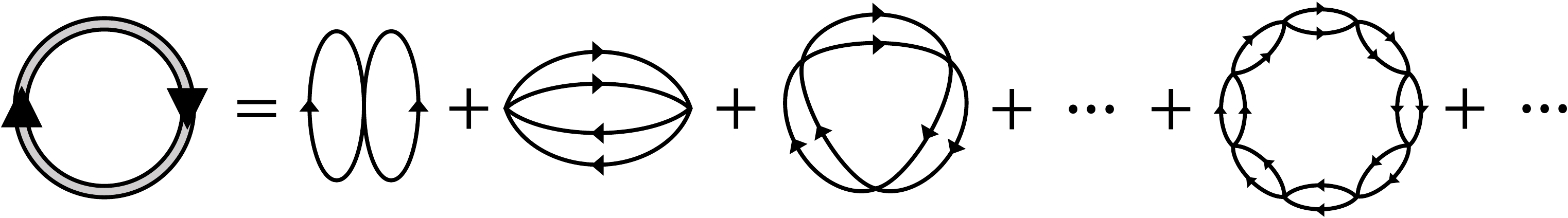}
\caption{Contribution of the diquark fluctuations to the thermodynamic potential.}
\label{fig_potential}
\end{figure}

\begin{figure}[tbp]
    \begin{tabular}{cccc}
      \begin{minipage}[t]{0.222\hsize}
        \centering
        \includegraphics[keepaspectratio, scale=0.1]{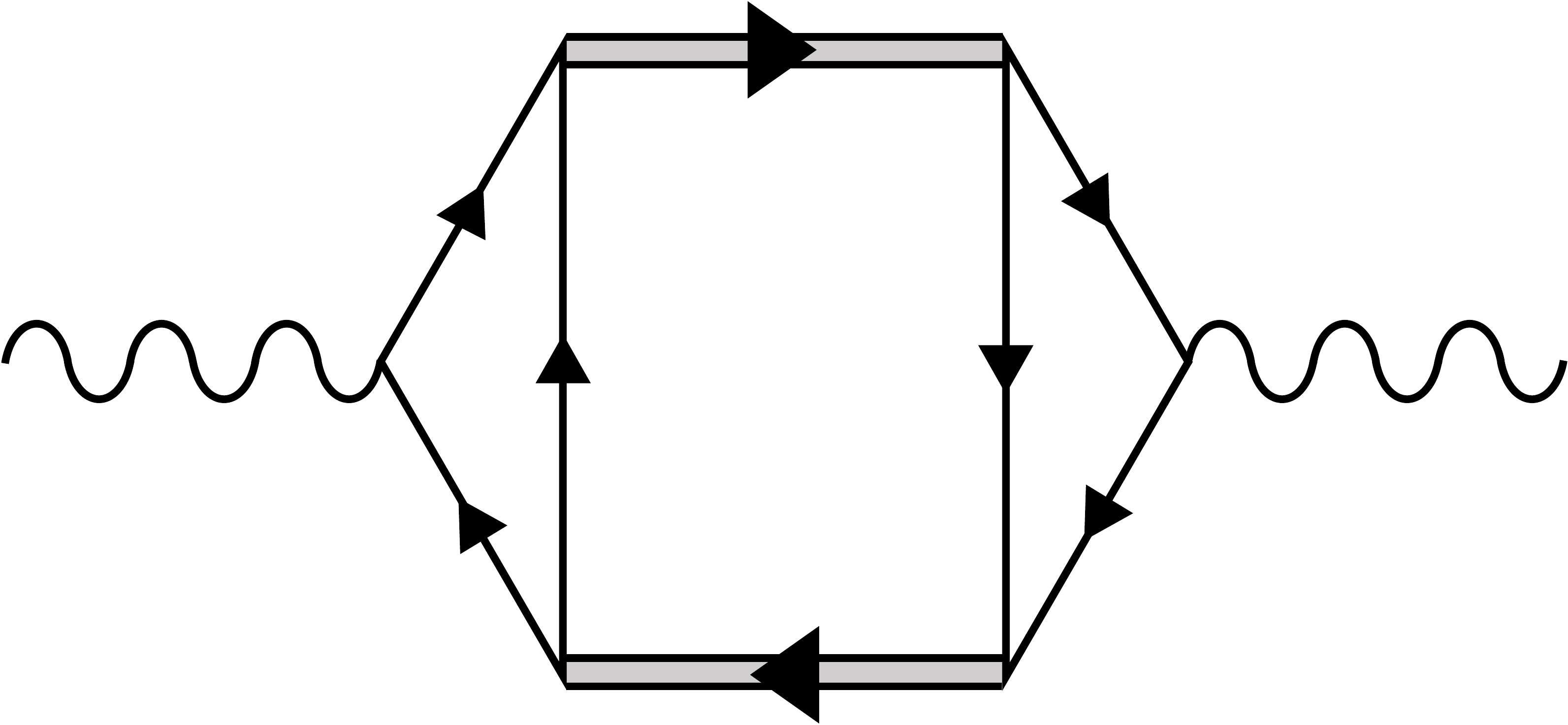}
        \subcaption{}
        \label{fig_AL}
      \end{minipage} &
      \begin{minipage}[t]{0.222\hsize}
        \centering
        \includegraphics[keepaspectratio, scale=0.1]{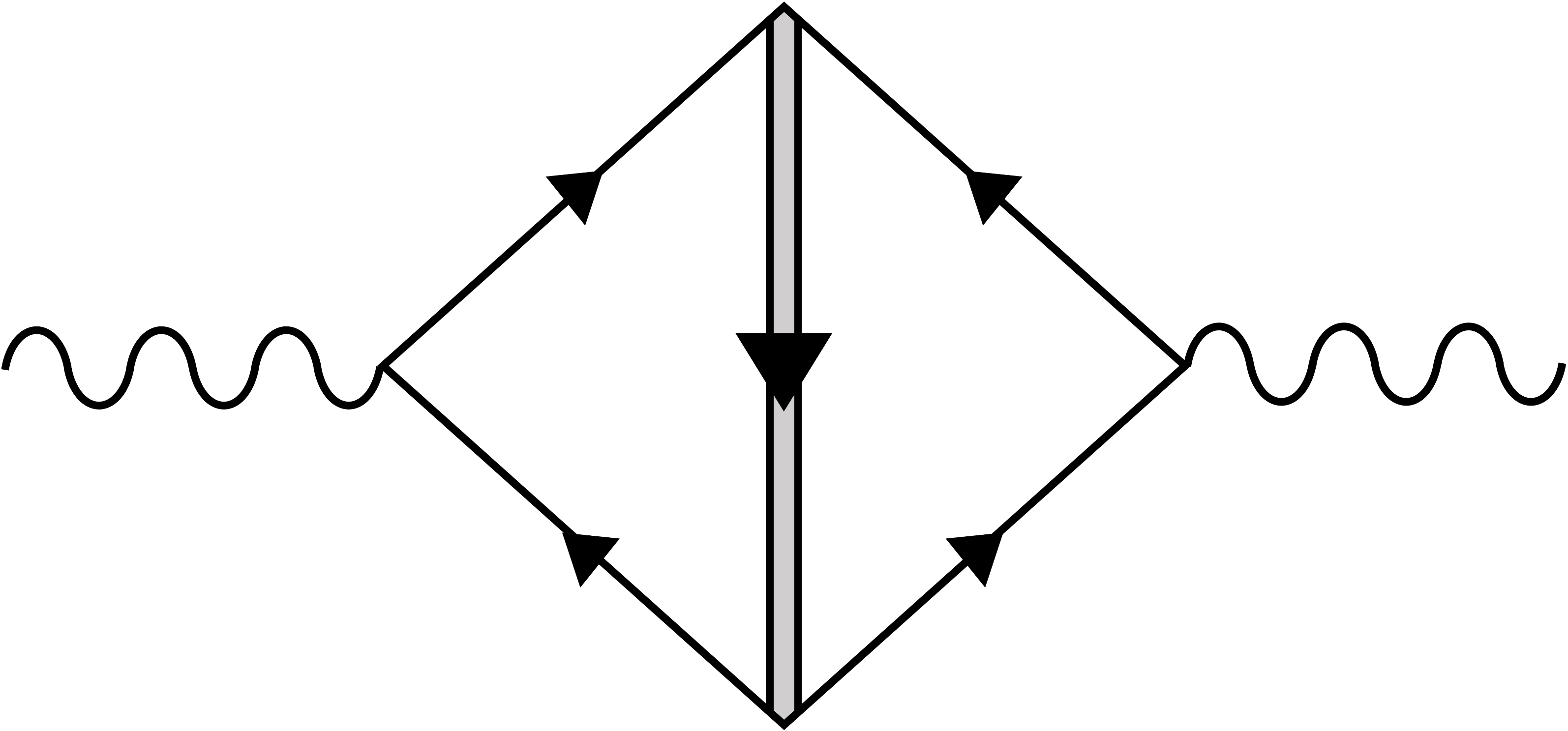}
       \subcaption{}
        \label{fig_MT}
      \end{minipage} &
      \begin{minipage}[t]{0.222\hsize}
        \centering
        \includegraphics[keepaspectratio, scale=0.1]{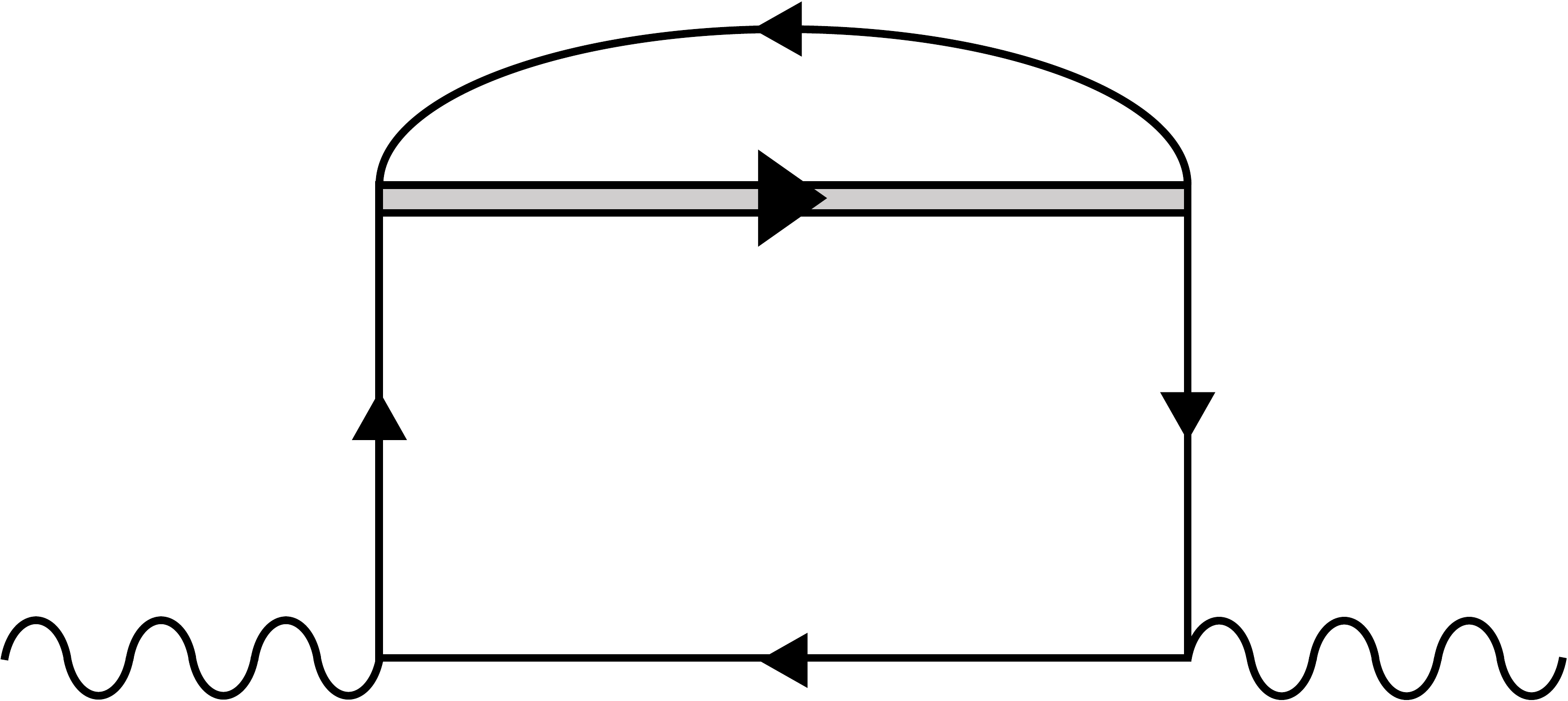}
        \subcaption{}
        \label{fig_DOS1}
      \end{minipage} &
      \begin{minipage}[t]{0.222\hsize}
        \centering
        \includegraphics[keepaspectratio, scale=0.1]{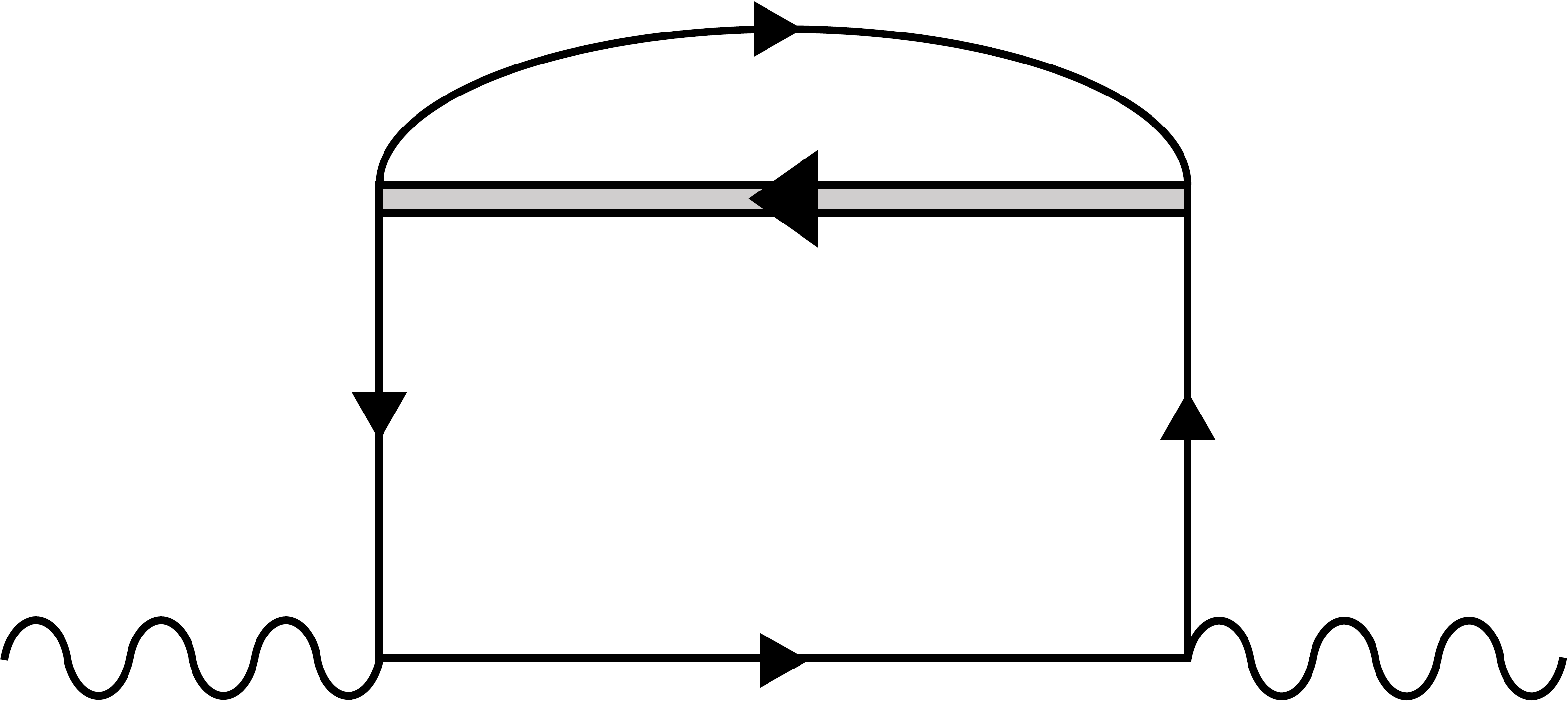}
        \subcaption{}
        \label{fig_DOS2}
      \end{minipage} 
    \end{tabular}
    \caption{Diagrammatic representations of the Aslamasov-Larkin (a),
      Maki-Thompson (b) and density of states (c,d) terms
      in Eqs.~(\ref{eq:AL})--(\ref{eq:DOS}).
      The wavy lines represent photons.}
     \label{fig_selfenergy}
\end{figure}

To incorporate the effects of the diquark fluctuations 
into the photon self-energy in a form that satisfies the WT identity,
we start from the one-loop diagram of $\Xi(\bm{q}, i\nu_n)$
shown in Fig.~\ref{fig_potential}, i.e. the lowest contribution of
diquark fluctuations to the thermodynamic potential.
The photon self-energy is then constructed by attaching 
electromagnetic vertices at two points of quark lines in Fig.~\ref{fig_potential}.
One then obtains four types of diagrams 
shown in Fig.~\ref{fig_selfenergy}.
These diagrams are called the Aslamasov-Larkin (AL) 
(Fig.~\ref{fig_selfenergy}(a))~\cite{ref_AL}, 
Maki-Thompson (MT) (Fig.~\ref{fig_selfenergy}(b))~\cite{ref_M}
and density of states (DOS) (Fig.~\ref{fig_selfenergy}(c, d))~\cite{book_Larkin} terms
in the theory of metallic superconductivity. 
Each contribution to the photon self-energy, 
$\Pi_{\rm AL}^{\mu\nu} (k)$, $\Pi_{\rm MT}^{\mu\nu} (k)$ and $\Pi_{\rm DOS}^{\mu\nu} (k)$, respectively, is given by
\begin {align}
\Pi_{\rm AL}^{\mu\nu} (k) &= 4 N_c \ T \sum_n \int \frac{d^3q}{(2\pi)^3} \Gamma^\mu(q, q+k) \Xi(q+k) \Gamma^\nu(q+k, q) \Xi(q), \label{eq:AL} \\
\Pi_{\rm MT}^{\mu\nu} (k) &= 
2 N_c \ T \sum_n \int \frac{d^3q}{(2\pi)^3} \Xi(q) R_{\rm MT}^{\mu\nu}(q, k), \label{eq:MT}  \\
\Pi_{\rm DOS}^{\mu\nu} (k) &= 2 N_c \ T \sum_n \int \frac{d^3q}{(2\pi)^3}
 \Xi(q) R_{\rm DOS}^{\mu\nu}(q, k),  \label{eq:DOS}
\end {align}
 with
\begin {align}
\Gamma^\mu (q, q+k) &= 2N_f (N_c-1) \ T \sum_m \int \frac{d^3p}{(2\pi)^3} 
{\rm Tr} [\mathcal{G}_0 (p)\gamma^\mu\mathcal{G}_0 (p+k)\mathcal{G}_0 (q-p)], \nonumber \\
R_{\rm MT}^{\mu\nu} (q, k) &= 2N_f (N_c-1) \ T \sum_m \int \frac{d^3p}{(2\pi)^3}
 {\rm Tr} [\mathcal{G}_0 (p)\gamma^\mu\mathcal{G}_0 (p+k)\mathcal{G}_0 (q-p-k)
\gamma^\nu\mathcal{G}_0 (q-p)], \nonumber \\
R_{\rm DOS}^{\mu\nu} (q, k) &= 2N_f (N_c-1) \ T \sum_m \int \frac{d^3p}{(2\pi)^3} 
\Big\{ {\rm Tr} [\mathcal{G}_0 (p)\gamma^\mu\mathcal{G}_0 (p+k)\gamma^\nu \mathcal{G}_0 (p) 
\mathcal{G}_0 (q-p)] \nonumber \\
&\qquad\qquad\qquad\qquad\qquad\qquad\quad + {\rm Tr} [\mathcal{G}_0 (p) \gamma^\mu
\mathcal{G}_0 (p-k) \gamma^\nu \mathcal{G}_0 (p) \mathcal{G}_0 (q-p)] \Big\}.
 \nonumber 
\end {align}
The total photon self-energy is then given by
\begin {align}
  \Pi^{\mu\nu} (k) = \Pi_{\rm free}^{\mu\nu} (k) + \Pi_{\rm fluc}^{\mu\nu} (k),
  \qquad
  \Pi_{\rm fluc}^{\mu\nu} (k) = \Pi_{\rm AL}^{\mu\nu} (k)
  +\Pi_{\rm MT}^{\mu\nu} (k)+\Pi_{\rm DOS}^{\mu\nu} (k) 
\label{eq_Pi}
\end {align}
where $\Pi^{\mu\nu}_{\rm free}(k)$ is the self-energy of the free quark 
system and $\Pi^{\mu\nu}_{\rm fluc}(k)$ denotes the modification of 
the self-energy due to the diquark fluctuations.
One can explicitly check that Eq.~(\ref{eq_Pi})
satisfies the WT identity.

\subsection{Time-dependent Ginzburg-Landau (TDGL) approximation}

The diagrams in Fig.~\ref{fig_selfenergy}
involve three-loop momentum integrals, which are 
cumbersome to compute. 
Therefore, we employ an approximation
that incorporates essential effects of the diquark fluctuations
near $T_c$ but, at the same time, allows us to evaluate the 
diagrams with a relative ease.

Since $[\Xi^R(\bm{0}, 0)]^{-1}=0$ at $T=T_c$ by the Thouless criterion, 
$\Xi^R(\bm{q}, \omega)$ in the low energy-momentum 
region may be well approximated near but above $T_c$ as follows,
\begin {align}
  \Xi^R (\bm{q}, \omega) = \frac{G_C}{c_0 \omega + c_1 \bm{q}^2 + a},
  \label{eq_softmodeapprox}
\end {align}
where the coefficients $c_0$, $c_1$ and $a$ can have $T$ dependence
and $a=0$ at $T=T_c$ from the Thouless criterion.
We determine these coefficients as $a = G_C [\Xi^R(\bm{0}, 0)]^{-1}$,
$c_0 = G_C (\partial/\partial \omega) [\Xi^R(\bm{0}, 0)]^{-1}$ and
$c_1 = G_C (\partial/\partial \bm{q}^2) [\Xi^R(\bm{0}, 0)]^{-1}$
from $\Xi^R(\bm{q}, \omega)$ obtained in the NJL model.
It is found that $c_0$ is complex while $c_1$ and $a$ real numbers.
The approximation Eq.~(\ref{eq_softmodeapprox}) is called the
time-dependent Ginzburg-Landau (TDGL) approximation in literature.
In Ref.~\cite{Kitazawa:2005vr}, it has been shown that 
Eq.~(\ref{eq_softmodeapprox}) reproduces the behavior of
$\Xi^R(\bm{q}, \omega)$ over wide ranges of $\omega$, $\bm{q}^2$ and $T(>T_c)$.

Next we consider similar approximations for the vertex functions
$\Gamma^\mu(q,q+k)$ and $R^{\mu\nu}(q,k) = R_{\rm MT}^{\mu\nu}(q,k)+R_{\rm DOS}^{\mu\nu}(q,k)$.
Here, we consider such approximations only for the spatial components
of these vertices because Eq.~(\ref{eq_RATE_kom}) can be obtained only
with the spatial components of $\Pi^{R\mu\nu}(k)$;
although Eq.~(\ref{eq_RATE_kom}) contains $\Pi^{00}(k)$,
this term is expressed in terms of the longitudinal part as
$\Pi^{R00} (k) = \bm{k}^2\Pi^{R11} (k)/k_0^2$ with $k=(k_0, |\bm{k}|, 0, 0)$
from the WT identity.
To approximate the spatial components $\Gamma^i(q,q+k)$ and $R^{ij}(q,k)$
to be consistent with Eq.~(\ref{eq_softmodeapprox}),
we substitute Eq.~(\ref{eq_softmodeapprox}) into
the WT identities for these vertices 
\begin {align}
&k_\mu \Gamma^\mu (q, q+k) = \Xi^{-1}(q+k)-\Xi^{-1}(q), \label{eq_ALvertex_Ward} \\
&k_\mu R^{\mu\nu} (q, k) = 2[\Gamma^\nu (q-k, q)-\Gamma^\nu (q, q+k)]. \label{eq_MTvertex_Ward}
\end {align}
Then, by comparing the lowest order terms of $\bm{q}$ and $\bm{k}$
in Eqs.~(\ref{eq_ALvertex_Ward}) and (\ref{eq_MTvertex_Ward})
we obtain
\begin {align}
\Gamma^i (q, q+k) = \frac{\partial \Xi^{-1} (q+k)}{\partial k_i} = 
\frac{c_1}{G_C} (2q+k)^i , \quad
R^{ij} (q, k) = 2 \frac{\partial \Gamma^j (q, q-k)}{\partial k_i} = 
- \frac{4 c_1}{G_C} \frac{k^i k^j}{\bm{k}^2}. \label{eq_vertex_approx} 
\end {align}
Each vertex in Eq.~(\ref{eq_vertex_approx}) is real, and this fact
simplifies the analytic continuation to obtain the retarded self-energy.
One also finds that the imaginary part of
$\Pi^{Rij}_{\rm MT}(q)+ \Pi^{Rij}_{\rm DOS}(q)$ calculated with
Eqs.~(\ref{eq_softmodeapprox}) and (\ref{eq_MTvertex_Ward})
vanishes. Therefore, the MT and DOS terms do not contribute
to the dilepton production rate.
This is in accordance with
the case of the metallic superconductivity~\cite{book_Larkin}.

\section{Numerical results of dilepton production rate}
\begin{figure}[tbp]
	\begin{tabular}{cc}
      \begin{minipage}[t]{0.48\hsize}
        \centering
        \includegraphics[keepaspectratio, scale=0.34]{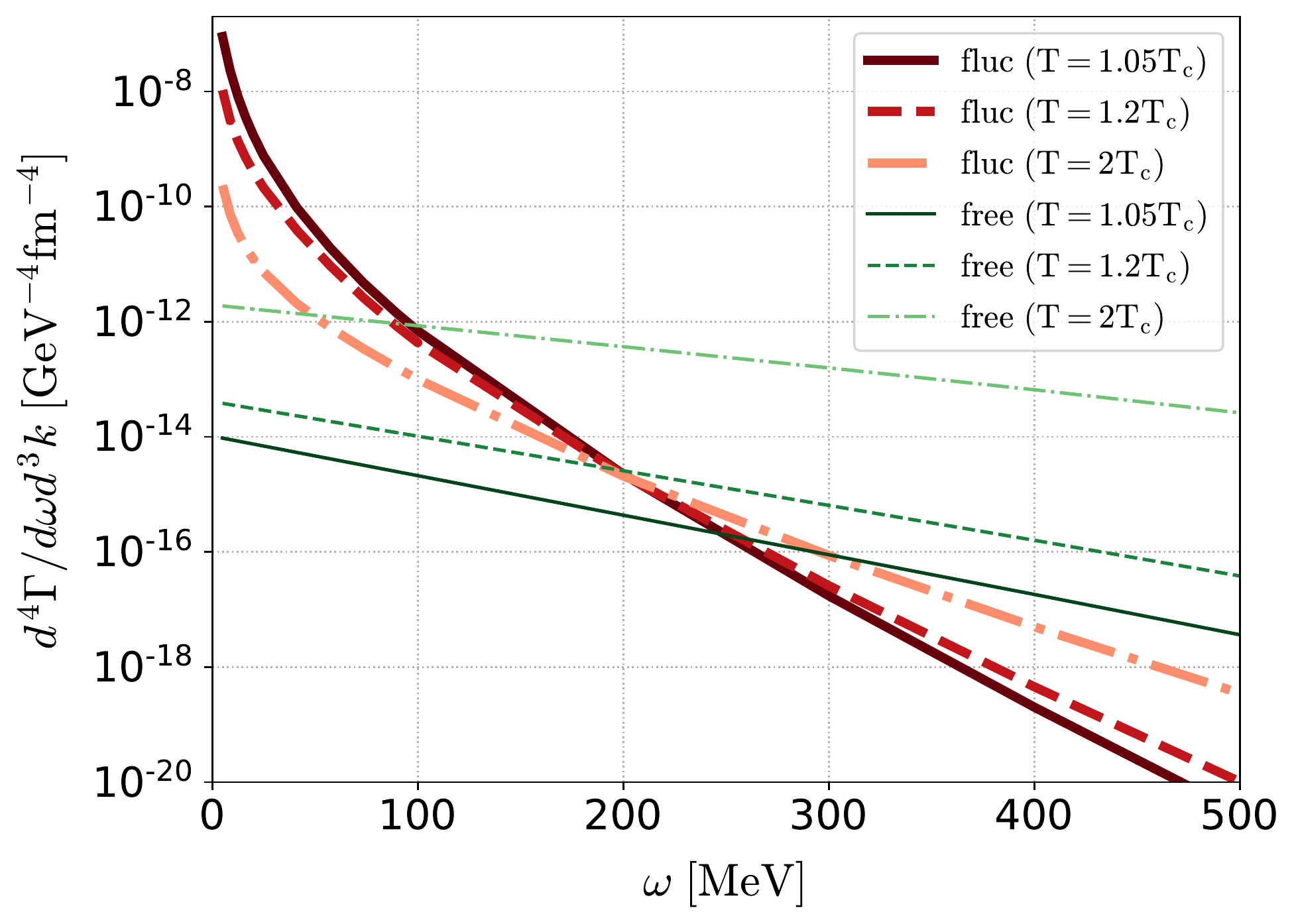}
      \end{minipage} &
      \begin{minipage}[t]{0.48\hsize}
        \centering
        \includegraphics[keepaspectratio, scale=0.34]{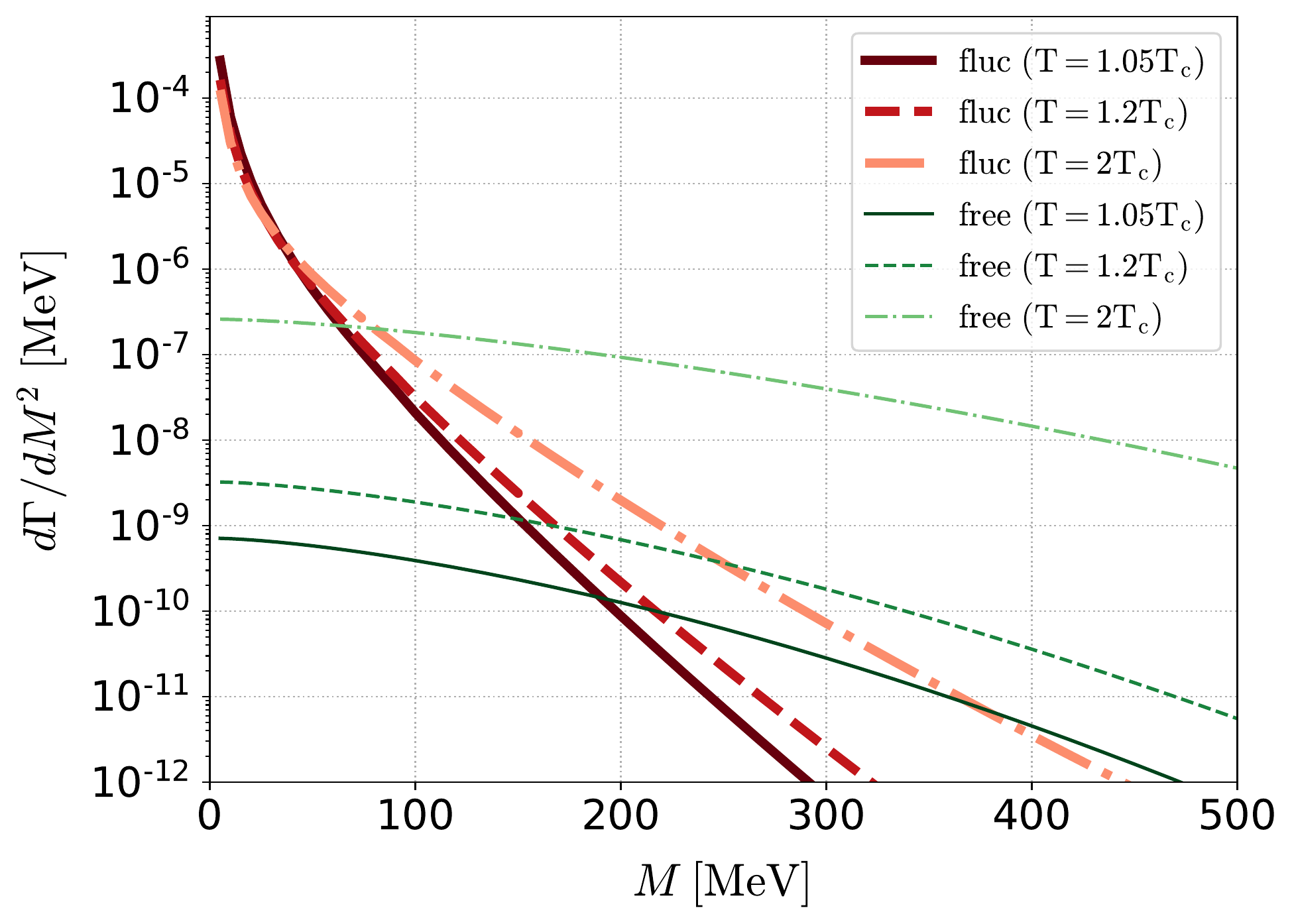}
      \end{minipage}
	\end{tabular}
        \caption{The dilepton production rates per unit energy $\omega$
          at $\bm{k}=0$ (left) and per invariant mass $M$ (right).
In each panel, the thick (red) lines are the contribution of 
diquark fluctuations $\Pi^{\mu\nu}_{\rm fluc}(k)$, and the thin (green) 
lines are the results for the free quark gas.
The solid, dashed and dash-dotted
lines are the results for $T=1.05 T_c$, $1.2 T_c$ and $2 T_c$.}
       \label{fig_RATE}
\end{figure}

In Fig.~\ref{fig_RATE}, we show the dilepton production rate
Eq.~(\ref{eq_RATE_kom}) calculated from the photon self-energy
Eq.~(\ref{eq_Pi}) at the quark chemical potential $\mu=350$~MeV
for several values of $T$ above $T_c$.
Shown in the left panel is 
the production rate per unit energy at $\bm{k}=0$.
The thick (red) lines show the contribution of diquark fluctuations
$\Pi_{\rm fluc}^{\mu\nu}(k)$, while the thin (green)
lines are the results for free quarks obtained from $\Pi_{\rm free}^{\mu\nu}(k)$.
The total rate is given by the sum of these two contributions.
The figure shows that the production rate from 
the diquark fluctuations is greatly enhanced 
in the low energy region compared with the free quark gas
for $T\lesssim2T_c$,
and this enhancement is more pronounced
as the system is closer to the critical temperature $T_c$.
This is not unexpected because the diquark 
fluctuations are the soft mode which acquires more concentrated
strength in the vicinity of the critical point.
In the right panel, we show the invariant-mass ($M$) spectrum 
\begin {align}
\frac{d\Gamma}{dM^2} = \frac{1}{2\omega} \int d^3k \frac{d^4\Gamma}{d^4k},
\end {align}
which is more convenient for a comparison with experimental data.
One sees that the enhancement of the production rate at small $M$
is observed in the invariant-mass spectrum for a similar temperature range,
while the $T$ dependence of the enhancement is milder than the left panel.

\section{Summary and concluding remarks}
In this study, we investigated the effect of diquark
fluctuations on the dilepton production rate
near but above the critical temperature of the 2SC.
The contribution of the diquark fluctuations was
taken into account so as to satisfy the WT identity
in the TDGL approximation.
It was found 
that the dilepton production rate
is greatly enhanced in comparison to 
the free-quark gas case 
in the low energy and low invariant-mass regions
near $T_c$.
This result suggests that such an enhancement 
can be used for the experimental signal for the existence of the
CSC phases.
In particular, the fact that the enhancement is seen even at $T\simeq2T_c$
would allow us to detect the signal in the HIC experiments
even when $T_c$ is so small
that the realization of the CSC phase itself in the HIC is impossible.

\end{document}